\documentclass[%
reprint,
showpacs,preprintnumbers,
showkeys,
nofootinbib,
 amsmath,amssymb,
 aps,
 pre,
 floatfix,
]{revtex4-1}

\usepackage[pdftex]{graphicx}
\usepackage{dcolumn}
\usepackage{bm}
\usepackage{hyperref}

\usepackage{microtype}

\begin{document}

\title{Temperature dependence of electrokinetic flux in Si nanochannel}

\author{B. Jelinek}
\affiliation{Center for Advanced Vehicular Systems,
  200 Research Boulevard,
  Starkville, Mississippi 39759
}

\author{S. D. Felicelli}
\email{felicelli@me.msstate.edu}
\affiliation{Mechanical Engineering Dept.,
  Mississippi State University,
  Mississippi State, Mississippi 39762}
\affiliation{Center for Advanced Vehicular Systems,
  200 Research Boulevard,
  Starkville, Mississippi 39759
}

\author{P. F. Mlakar}
\author{J. F. Peters}
\affiliation{U.S.\ Army ERDC,
  3909 Halls Ferry Rd,
  Vicksburg, Mississippi 39180
}

\date{\today}

\begin{abstract}
  Significant temperature effects on the electrokinetic transport in
  a nanochannel with a slab geometry are demonstrated using a
  molecular dynamics (MD) model. A system consisting
  of Na$^{+}$ and Cl$^-$ ions dissolved in water and confined between
  fixed crystalline silicon walls with negatively charged inner
  surfaces in an external electric field was investigated.
  Lennard-Jones (LJ) force fields and Coulomb electrostatic
  interactions with Simple Point Charge Extended (SPC/E) model were
  used to represent the interactions between ions, water molecules,
  and channel wall atoms.
  Dependence
  of the flow of water and ions on the temperature was examined. The magnitude of the water flux
  and even its direction are shown to 
  be significantly affected by temperature.
  In particular, the previously reported flow reversal phenomenon
  does not occur at higher temperature.
  Temperature dependence of the flux was attributed to the charge
  redistribution and to the changes in viscosity of water.
\end{abstract}



\maketitle    
\section{Introduction}

Nanoscale numerical models of electro-osmosis~\cite{freund:2002eo,qiao2004inv,JosephS._la0610147,Rotenberg07:multisc_ion_diff_clays,huang2007:ion_spec,lorenz2008molecular}
provide insight into the important transport
mechanism, help improve current technological devices, and guide the
design of new technology based on the principles of electrokinetic
transport.
They also improve understanding of transport processes in biology
\cite{Han2006285,*Cory2007:interf,*Gumbart20091129,*cruz2009ionic,*Lin20101009}
and their temperature dependence~\cite{Jones2009354}.

The present study examines the fluid flow and ion species
transport under an electric field in nanochannels typically found in
heterogeneous porous media. Although the continuum conservation
equations can be applied to microscale channels, the main driving
force of electrokinetic transport occurs in an electric double layer
(EDL) at the solid liquid interface with dimensions that can be
comparable to intermolecular distances. Therefore,
molecular dynamics (MD) 
simulations were applied to analyze the interaction between ions, water
molecules, and wall atoms in the EDL region.
Time averaged velocity and concentration profiles of
water molecules and ionic species at different temperatures were obtained from these MD simulations.
Water viscosity profile across the channel was estimated at different
temperatures and its impact on the transport of water and ionic
species will be discussed.

\section{MD potentials}

\begin{video}[t]
\includegraphics[width=1.0\columnwidth]{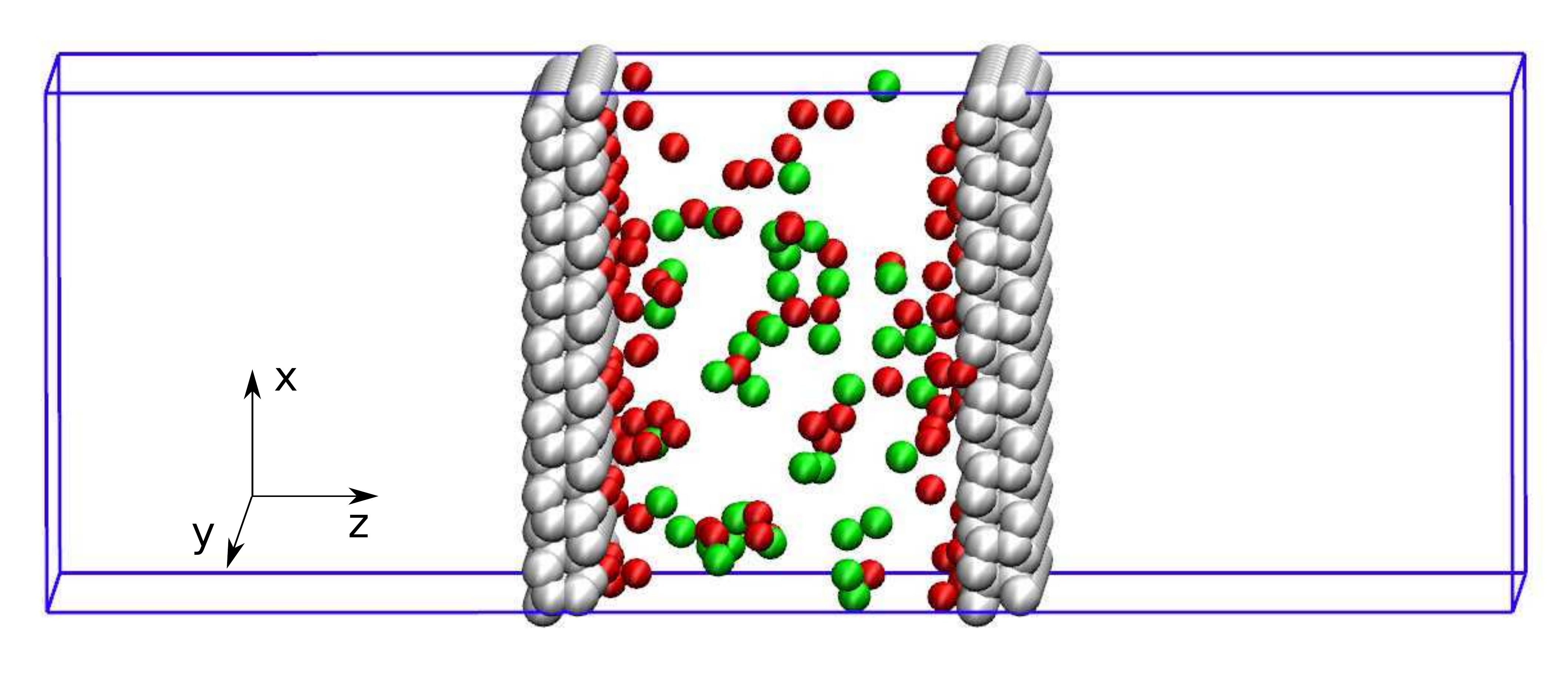}
\setfloatlink{http://www.msstate.edu/~bj48/untitled.mpg}
\caption{(Color online) \label{vid:box}Simulation box without water molecules.
  Si wall atoms are gray, Na$^+$ ions red,
  and Cl$^-$ ions green. Water molecules are not shown.
  Accompanying animation shows movement of ions when electric field is applied in
  the positive x direction.}
\end{video}

Models based on classical
Lennard-Jones (LJ) force fields and Coulomb electrostatic interactions
with Simple Point Charge Extended (SPC/E)~\cite{Berendsen1987} model were used to
represent the interactions between ions, water molecules, and channel
wall atoms.
LJ contribution of atoms $i$ and $j$ to the total potential energy is
\begin{equation}
V_{LJ}(r) =  
4\epsilon_{ij}
\left[
\left( \frac{\sigma_{ij}}{r} \right)^{12} -
\left( \frac{\sigma_{ij}}{r} \right)^{6}
\right].
\label{eqn:lj}
\end{equation}
The parameters $\epsilon_{ij}$ (the depth of the potential well) and $\sigma_{ij}$ ($\sqrt[6]{2}\sigma_{ij}$ is the minimum energy distance) depend on atomic
species of $i$th and $j$th atoms (Table~\ref{tab:lj}).
The Coulomb electrostatic potential energy contribution is
\begin{equation}
V_C(r) = \frac{1}{4\pi\varepsilon_0}\frac{q_i q_j}{r}.
\label{eqn:coul}
\end{equation}
The $\varepsilon_0$ is vacuum permittivity, $q_i$ and $q_j$ are
charges of $i$th and $j$th atoms, and $r$ is the distance
between $i$th and $j$th atoms.
The Particle-Particle Particle-Mesh (PPPM)~\cite{hockney:comp_sim_part,*plimpton97:pppm} method
was used for long range electrostatics.

\begin{table}[t]
\caption{\label{tab:lj}Parameters of Lennard-Jones potentials (from the \textsc{gromacs}~\cite{berendsen1995gromacs,*lindahl2001gpm,*Hess08_gromacs4,*patra2004systematic} force field), $\sigma_{ij}$ in $\text{\AA}$, $\epsilon_{ij}$ in $\text{cal/mol}$.}

\begin{tabular}{c|c|c|c|c|c|c|c|c|c|c}
ij       & OO  & OSi & ONa & OCl & SiSi&SiNa&SiCl&NaNa& NaCl& ClCl\\ 
\hline                                                                       
$\sigma_{ij}$  & 3.17 & 3.27 & 2.86 & 3.75 & 3.39 &2.95 &3.88 &2.58 & 3.38 & 4.45 \\
$\epsilon_{ij}$& 155  & 301  & 47.9  & 129 & 584  &92.9 &249  &14.8 & 39.6 & 106\\
\hline
\end{tabular}
\end{table}

\section{Simulation setup}

In order to study the temperature effects, the authors first reproduced the system
studied previously \cite{qiao2004inv}, a summary of which is given next.
The dimensions of the solution region were 4.66x4.22x3.49 nm.
Channel walls, perpendicular to the z axis, were formed by four
[111] oriented layers of Si atoms in a diamond crystal structure,
each wall being 0.39 nm thick.
Periodic boundary conditions were applied in the x and y directions.
The size of the simulation cell in the z direction was extended
to three times outermost-to-outermost wall layer distance (3x4.37~nm)
to mitigate electrostatic interactions of periodic images in the z direction.
The electric field of 0.55~V/nm was applied in the positive x direction.

The crystalline channel walls consisted of 1232 fixed Si atoms.
A charge of $-70$ e was distributed uniformly on atoms
of the innermost surface layers of both Si walls
(total 308 Si atoms were charged negatively and 924 were uncharged).
Therefore, the innermost wall surface atoms had $-0.227273$ e/atom charge.
That corresponds to a surface charge density of $-0.285$~C/m$^2$, which is 
close to typical value for a fully ionized surface $0.2$~C/m$^2$
\cite{israelachvili1985:intermolecular} or the charge density of $-0.1$
C/m$^2$ measured at silica surface~\cite{dove2005surfchrg}.
Then 2290 water molecules were inserted avoiding close contacts,
and randomly selected 146 water molecules were
replaced by 108 Na$^+$ and 38 Cl$^-$ ions.
Thus the whole system was electrically neutral and resulting NaCl concentration
in the channel center at the temperature of 300 K was $\approx$1.2~M.

First, the energy minimization
of the system was performed using the conjugate gradient method. Then, the
system was equilibrated by 2 ns of MD simulation without an electric field.
A timestep of 2x10$^{-15}$ s was used
for the leapfrog~\cite{rap2004:the_art} integration of Newton's equations of motion.
The resulting configuration, excluding water molecules, is shown in Video~\ref{vid:box}.
Finally, a 22 ns MD run was performed
with external electric field.
The SHAKE~\cite{ryckaert77shake} algorithm was used to constrain bonds
of water molecules. 
The solution temperature was controlled by the Nos\'{e}-Hoover
\cite{nose84:unif_form,*hoover85:can_dyn} selective thermostat
(Sec.~\ref{sec:thermostat}).


\section{Thermostat\label{sec:thermostat}}

\begin{figure}[h]
\includegraphics[]{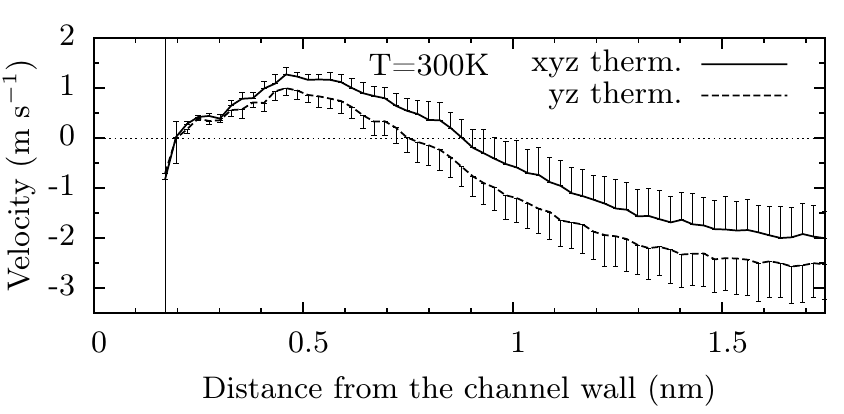}
\caption{{} Comparison of water velocity profiles using
  full thermostat (adjusting x, y, and z velocity components) and
  selective thermostat (adjusting y and z components only).
  Error bars are only shown on one side of the values
  to avoid overlap. The values shown (on all figures) are averages
  over $\approx$0.026 nm wide bins parallel to the xy plane that are then
  symmetrically averaged about the channel center. \label{fig:therm}}
\end{figure}

The thermostat in MD simulations adjusts the velocity to
maintain the desired temperature and its proper fluctuations.
To check how much the velocity component in the field direction is
affected by thermostat, the
velocity profiles from simulation with selective thermostat (adjusting y and z velocity
components only) were compared to results with full thermostat (adjusting x, y, and z
components of velocity). The number of thermal degrees of freedom
was adjusted accordingly. Figure~\ref{fig:therm} shows that selective
thermostat produces slightly more negative velocity.

Velocity profiles with profile-unbiased
thermostat (PUT) \cite{evans86:put} that ``preserves'' the 
velocity profile (i.e.\ its x component) across the channel were also generated.
Averages did not differ significantly 
from those obtained by selective thermostat (adjusting only y and z
velocity components), but PUT error bars were larger, therefore 
the selective thermostat (adjusting y and z components of velocity only) was used.

The authors then verified that velocity profiles with full thermostat obtained from
\textsc{gromacs}~\cite{berendsen1995gromacs,*lindahl2001gpm,*Hess08_gromacs4}
were statistically identical to those from \textsc{lammps}~\cite{plimpton95:fpa} when
thermostating all velocity components, even though \textsc{gromacs} used the
PME (Particle-Mesh-Ewald) method \cite{essmann1995smooth} with a slab
correction \cite{yeh1999ewald} in the z direction for long range
electrostatics and SETTLE~\cite{miyamoto1992settle} algorithm to
constrain bonds of water molecules.

There is a noticeable difference between the velocity profile
in Fig.~\ref{fig:therm} and the one
of~\cite{qiao2004inv}. The velocities from the present work are more positive.
That is consistent with the positive peak in the driving force 0.65 nm from
the channel wall in Fig.~\ref{fig:df}, whereas the
driving force of~\cite{qiao2004inv} remain negative in
that region. Assuming that all the parameters were set correctly, the
difference is hard to track since the
\textsc{gromacs}~\cite{berendsen1995gromacs,*lindahl2001gpm,*Hess08_gromacs4}
package used by~\cite{qiao2004inv} does not offer a selective
thermostat.

\section{Concentration profiles}

\begin{figure}[b]
\includegraphics[]{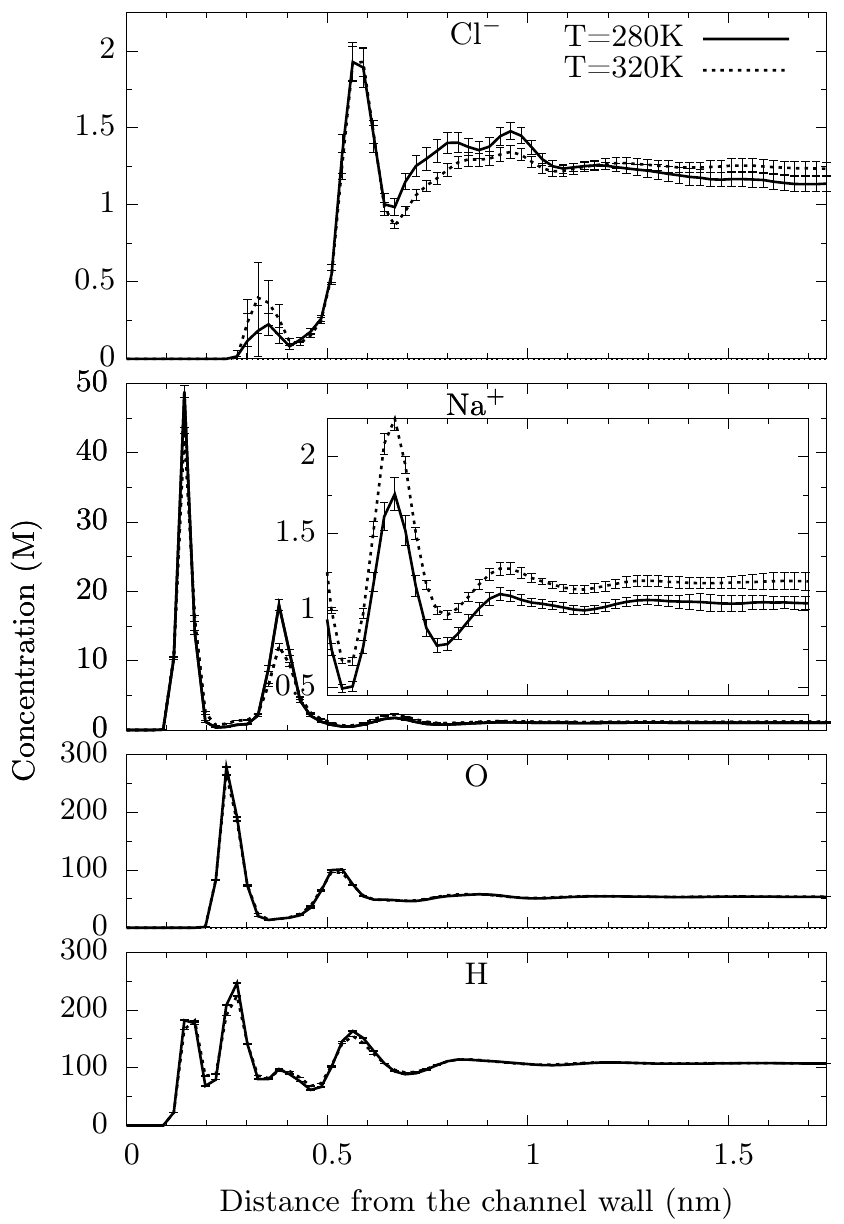}
\caption{{} Concentrations of Na$^+$ and Cl$^-$ ions, and
  H and O atoms from H$_2$O molecules symmetrically averaged across
  half of the channel. The inside plot is a magnified version of a
  larger plot. Concentration of Na$^+$ ions in the channel center
  increases as the temperature increases. \label{fig:conc_nacl}}
\end{figure}

The calculated atomic concentration profiles shown in Fig.~\ref{fig:conc_nacl},
which agree well with reported results \cite{qiao2004inv},
manifest formation of alternatively charged layers of atoms parallel
to the channel walls.
The negatively charged Si wall attracts both positively charged Na$^+$
ions and slightly positively charged H atoms (0.4238 e/atom for SPC/E
water) from H$_2$O molecules, forming the first near-wall
concentration peak.
The adjacent layer is formed of slightly negatively charged O atoms
($-0.8476$ e/atom for SPC/E water) from water molecules.
Five noticeable layers with alternating charge signs are formed---four ionic layers can be seen in Fig.~\ref{fig:df} and
the fifth is a negatively charged layer of O atoms
0.25 nm from the channel wall (Fig.~\ref{fig:conc_nacl}).
Layering of particles near a flat surface is characteristic for polar
liquids~\cite{Bressanini1992:charge_lay}, LJ
fluids~\cite{abraham1978:interfacial}, and charged
surfaces~\cite{lyklema1998:electrokinetics,*netz2003:electrof}.
Further towards the channel center, the concentrations of ions are more balanced ($\approx$1.2~M).


\section{Driving force\label{sec:df}}

\begin{figure}[!b]
\includegraphics[]{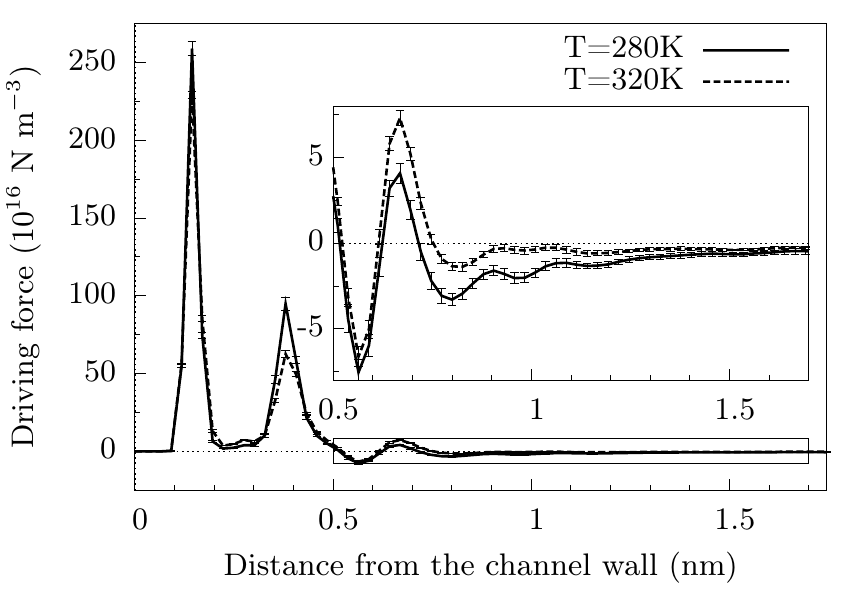}
\caption{{} Driving force for the fluid flow given by Eq.~(\ref{eq:fd}). Values are symmetrically averaged across half of the channel. Inside plot magnifies values in the central region of the channel---where driving force becomes more positive at higher temperature. The temperature changes in concentration profiles, resulting in changes of driving force, are the main factor contributing to the temperature dependence of the flow. \label{fig:df}}
\end{figure}

Since the net charge of a standalone water molecule is zero, its center of mass
will not move in the presence of external electric field.
The driving force for the electro-osmotic flow comes from the electric field
causing the movement of a fluid region with non-zero net charge.
Regions of fluid with positive net charge will drive the flow in the
direction of an external electric field, while the regions with
negative net charge will drive the flow in the direction opposite to
an external electric field.
The driving force is defined as
\begin{equation}
{\bf F}_d(z)=e\left[c_{\text{Na}^+}(z)-c_{\text{Cl}^-}(z)\right]{\bf E}_{ext},
\label{eq:fd}
\end{equation}
where $e$ is the elementary charge, $c_{\text{Na}^+}(z)$ and $c_{\text{Cl}^-}(z)$ are 
ionic number densities across the channel, and ${\bf E}_{ext}$ is an external
electric field.
Figure~\ref{fig:df} shows a dependence of the driving force on temperature.

When the temperature is increased, the driving force in the region
further than 0.42~nm from the channel wall becomes more positive
(because of increased Na$^+$ concentration in that region), and water
starts to flow in the positive x direction.  The driving force from
Cl$^-$ ions, on the contrary, will increase only in the near-wall region,
as some of the Cl$^-$ ions will redistribute closer than 0.5~nm within the
channel wall. In the following section it will be argued that this charge
redistribution at increased temperature drives the flow
in the positive x direction.

\begin{figure*}[!t]
\includegraphics[]{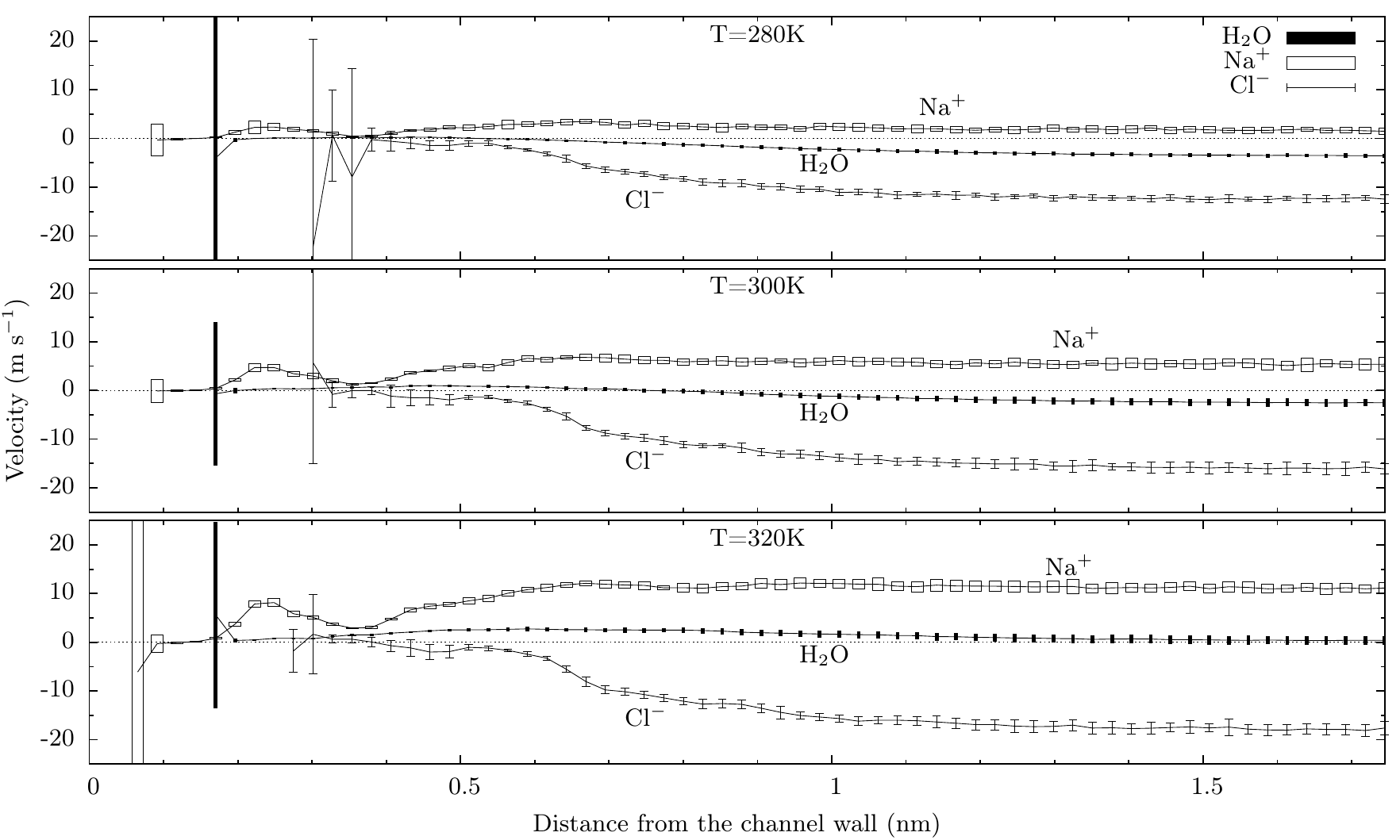}
\caption{{} \label{fig:temp_vel}Velocity profiles from MD
  at T=280 K (top), 300 K, and 320 K (bottom).  Negative flow means
  movement in the direction opposite to the applied electric field.
  Velocity of H$_2$O is a velocity of oxygen atoms from H$_2$O
  molecules.  Note that the temperature increase changes the water
  flow direction along the channel.  Error bars (for Na$^+$ and H$_2$O
  they are represented by rectangles) were obtained by analysis of ten
  simulations with different seeds of initial Gaussian random velocity
  distribution.
  Large error bars near the wall are due to low concentrations
  (Fig.~\ref{fig:conc_nacl}).
  Details of water velocity profiles are shown in
  Fig.~\ref{fig:3harm}.}
\end{figure*}

\section{Temperature dependence of velocity profiles\label{sec:tdep}}

A significant dependence of the flow on temperature was observed.
Figure~\ref{fig:temp_vel} shows the dependence
of the velocity profile of ions and water on the
temperature.
It demonstrates that the water flow and even its direction are
directly affected by the temperature. At lower temperatures, flow
reversal of water was observed, in agreement with previous results
\cite{qiao2004inv}. However, at higher temperatures, the water flows
in the direction expected by a standard EDL model.

Note that even though Cl$^-$ ions in water move faster than Na$^+$
ions (Fig.~\ref{fig:temp_vel}), that does not affect water velocity
profile. Water movement is not governed by the ionic velocities, but
by the profile of the driving force. Velocity of ions relative to
water is given by their respective mobilities $\mu_i$ according to the
definition of mobility ${\bf v_i}=\mu_i {\bf E}$. The LJ potential
properly reflects the experimental mobility of Cl$^-$ ions being
higher than the mobility of Na$^+$ ions. Mobility of ions will
affect the rate of ion accumulation on electrodes in an experiment,
but not the velocity profile of water in the present model.

The temperature dependence of the electro-osmotic flow was attributed
to (a) charge redistribution and to (b) changes of water viscosity
with temperature.

As mentioned in
Sec.~\ref{sec:df}, the positive ions will redistribute towards the
channel center at higher temperatures
(Fig.~\ref{fig:conc_nacl}--\ref{fig:df}). At T=320 K, Na$^+$ ions
dominate Cl$^-$ ions in the region 0.61 to 0.76 nm from the
wall. The increased Na$^+$ concentration in the channel center comes
from lowering two near-wall
concentration peaks of Na$^+$ ions (at 0.14 and 0.38 nm from the wall in
Fig.~\ref{fig:conc_nacl}), that redistribute further than 0.42 nm
from the channel wall at higher temperature.
This increased Na$^+$ concentration in the channel center (where the
water viscosity is lower than at the surface---see Sec.~\ref{sec:visc})
stimulates the movement of water in the positive x direction.
Even though Na$^+$ ions dominate only in regions less than 0.52 nm
and 0.61 to 0.76 nm from the wall,
hydrogen bonding and collisions will also drag adjacent layers of water
in the channel center, outperforming the competing mechanism of water
dragged in the negative x direction by Cl$^-$ ions that have higher
concentration than the Na$^+$ in the channel center.

On the contrary, at lower temperatures Cl$^-$ ions are more
dominant over Na$^+$ ions in the channel center, causing a flow in the
negative x direction.

\section{Relation of velocity profiles and charge distribution\label{sec:stok}}

Velocity profile is related to charge density profile
by the Stokes equation~\cite{hunter1981zpc,*hunter1987foundations}
\begin{equation}
\frac{d}{d z} \left[\eta(z) \frac{d u_x(z)}{d z}\right] = -F_d(z).
\label{eq:stokes}
\end{equation}
The magnitude of the driving force, $F_d(z)$, given by
Eq.~(\ref{eq:fd}),
as well as the velocity profile, $u_x(z)$, can
be calculated as averages from MD simulations.
The $\eta(z)$, viscosity profile of water across the channel, can then
be estimated from MD as follows.

\section{Estimation of water viscosity\label{sec:visc}}

The water viscosity profile across the
channel was estimated following the method of~\cite{freund:2002eo}. 
The same method (except smoothing of velocity profile) was used
by~\cite{qiao2003ion}.
To simplify equations, the coordinate system with z=0 at
the channel center will be used. Integration
of Stokes Eq.~(\ref{eq:stokes}) leads to viscosity estimate
\begin{equation}
  \left.\eta(z)\right|_{z=z_0} = 
  \frac
  {-\displaystyle\int_{0}^{z_0} \! F_d(z)\,dz}
  {\left.\dfrac{du_x(z)}{dz}\right|_{z=z_0}}.
  \label{eq:visc}
\end{equation}
The fit diverges near the points where the derivative of velocity is
zero. This is appropriate, since the viscosity can not be estimated in
the region with zero shear strain---which is the term in denominator. 
The numerator term represents the shear stress~\cite{qiao2003ion}.

To obtain smooth derivative of $u_x(z)$, the velocity profile was
approximated by the sum of harmonic components
\begin{equation}
u_{xfit}\left(z\right) = \sum_{n=0}^7 a_n
\cos
\left(n\pi
\frac
{z}
{h}
\right).
\end{equation}
The $h$ is the distance of the furthest oxygen atom from the channel
center. In contrast to the original
approximation of~\cite{freund:2002eo}, the exponential term is
excluded. Also, to exploit the symmetry of the system, the origin of
cosine components is set to the channel center.

\begin{figure}[!t]
\includegraphics[]{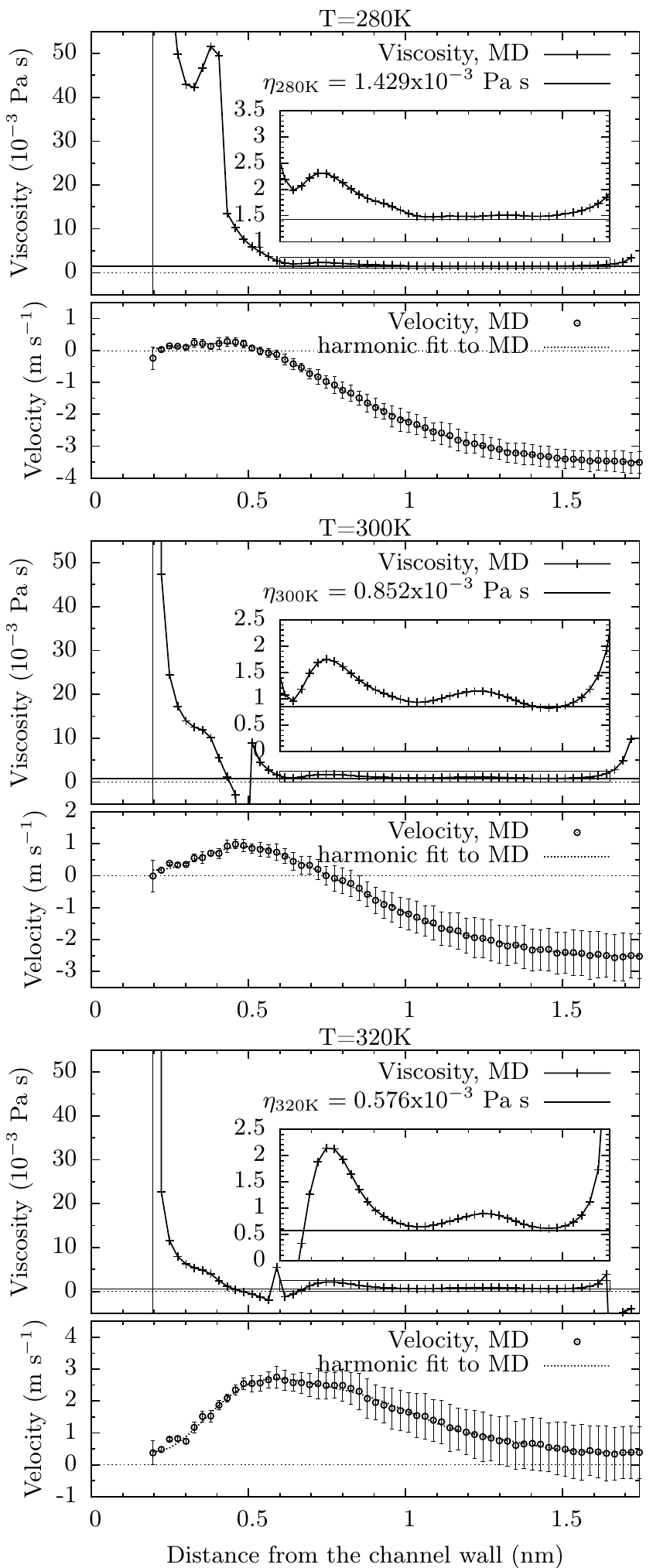}
\caption{{} Velocity profile fit to the sum of harmonics
  (lower plots) and viscosity estimate from
  Eq.~(\ref{eq:visc}) (upper plots) at T=280 K, 300 K and 320 K. The inside plots are magnified versions of larger plots.
  \label{fig:3harm}}
\end{figure}

The velocity, its approximation, the viscosity profile estimated by
Eq.~(\ref{eq:visc}), and the experimental viscosity of
water~\cite{site:visc} at the simulated temperatures are shown in
Fig.~\ref{fig:3harm}. Note that the estimated
viscosity (the line made of $+$ symbols on the upper plots in
Fig.~\ref{fig:3harm}) at its minimum reproduces
the experimental viscosity (solid horizontal line).

At lower temperatures, the viscosity of near-wall water layers
increases drastically. That means the near-wall Na$^+$ ions can not
drive much water flow. Conversely, at higher temperatures, the
near-wall water will become more mobile, exhibiting a partial
slip at T=320 K.

\section{Conclusions}

Many technological processes and phenomena in live organisms exhibit a
strong dependence on temperature. It is important to understand how
the temperature affects technological and natural processes.
This work indicates that a variation of temperature
can have deterministic effects on an electrokinetic flow.

This paper revisited the phenomenon of flow reversal during
electrokinetic flow in a slit nanochannel. It was found, using a
molecular dynamics analysis, that both the magnitude and direction of
the electro-osmotic flow are significantly affected by temperature.
Even though the results reported
in~\cite{qiao2004inv} could be replicated, it was found that the flow reversal of water
molecules no longer occurs if the temperature
is increased slightly above the 300~K used in~\cite{qiao2004inv}.
The mechanisms that lead to such a significant
temperature dependence of the nanochannel flow were then investigated. In particular, it was
shown that the temperature dependence of the flux can be explained by
the charge redistribution and the decrease of near-wall water
viscosity at higher temperatures.

\section*{Acknowledgments}
This work was funded by the US Army Corps of Engineers through
contract number W912HZ-09-C-0024. Computational resources at the MSU
HPC$^2$ center were used. 
Computational packages \textsc{lammps}~\cite{plimpton95:fpa} and
\textsc{gromacs}~\cite{berendsen1995gromacs,*lindahl2001gpm,*Hess08_gromacs4}
versions 4.0.4 were used to perform MD simulations.
Video~\ref{vid:box} was made using Visual Molecular
Dynamics~\cite{humprey96:vmd} package.  Part of this work was
disclosed in~\cite{jelinek2009asme}.  Permission to publish this
material was granted by the Director of the Geotechnical and
Structures Laboratory of the Engineer Research and Development Center,
U.S.\ Army Corps of Engineers.

\end{document}